\documentclass[manuscript]{aastex}

\shorttitle{ALFV\'EN WAVES, DAMPING AND FLARES}
\shortauthors{RUSSELL \& FLETCHER}

\begin{document}


\title{PROPAGATION OF ALFV\'ENIC WAVES FROM CORONA TO CHROMOSPHERE AND CONSEQUENCES FOR SOLAR FLARES}


\author{A. J. B. Russell \and L. Fletcher}
\affil{SUPA School of Physics \& Astronomy, University of Glasgow, Glasgow, Scotland, U.K.}


\begin{abstract}
How do magnetohydrodynamic waves travel from the fully ionized corona, into and through the underlying partially ionized chromosphere,
and what are the consequences for solar flares?  
To address these questions, we have developed a 2-fluid model (of plasma and neutrals)
and used it to perform 1D simulations of Alfv\'en waves in a solar atmosphere with realistic density and temperature structure.  
Studies of a range of solar features (faculae, plage, penumbra and umbra)
show that energy transmission from corona to chromosphere can exceed 20\% of incident energy
for wave periods of one second or less.
Damping of waves in the chromosphere depends strongly on wave frequency:
waves with periods 10 seconds or longer pass through the chromosphere with relatively little damping,
however, for periods of 1 second or less, a substantial fraction (37\%--100\%) of wave energy entering the chromosphere is damped by 
ion-neutral friction in the mid and upper chromosphere, with electron resistivity playing 
some role in the lower chromosphere and in umbras.
We therefore conclude that Alfv\'enic waves with periods of a few seconds 
or less are capable of heating the chromosphere during solar flares, 
and speculate that they could also contribute to electron acceleration or exciting sunquakes.
\end{abstract}



\section{Introduction}
During solar flares, up to $10^{26}\mbox{ J}$ of energy is released in the corona by reconfiguration of magnetic fields
and a significant fraction of this energy is transported $\sim10^5\mbox{ km}$ to the chromosphere
to power the UV and white light emissions that make up the majority of flare electromagnetic radiation.
This energy transport is widely ascribed to beams of electrons
that are assumed to be accelerated in the corona in the vicinity of the magnetic energy release.
Such models are appealing because electron beams impacting on the chromosphere can explain 
the hard x-ray (HXR) sources associated with flares, 
using the collisional thick target model \citep{1971Brown,1972Hudson}.

Electron-beam-only models face a number of theoretical difficulties which have been described elsewhere, 
e.g. \citet{1990Brown,2008FletcherHudson,2009Brown}.
For example, coronal acceleration models require large numbers of particles when compared to typical coronal densities,
or else a powerful resupply mechanism which is opposed by microinstabilities.
Further difficulties include the small area of HXR footpoint sources determined by RHESSI observations,
which imply electron beam densities in the corona high enough for the beams to be unstable \citep{1977BrownMelrose,2011Krucker}.
It is therefore worthwhile to consider other mechanisms by which at least some portion of flare energy could be transported
to the chromosphere and deposited there.

An alternative transport mechanism that has lately received renewed attention is magnetohydrodynamic (MHD) waves.
Recent advances in observations have led to a consensus that the solar atmosphere is pervaded by MHD waves,
which are now observed almost routinely
\citep[e.g.][and references therein]{1997Ofman,1998Deforest,1999Thompson,1999Robbrecht,
1999Aschwanden,1999Nakariakov,2007DePontieu,2007Okamoto,2007Tomczyk,2009Jess,2011McIntosh}.
Against this background, it has been proposed that a substantial part of the energy released by a flare is in the form of
MHD waves, providing an alternative mode of energy transport to the chromosphere \citep{2008FletcherHudson},
possibly operating alongside electron beams.

There are a number of theoretical and observational arguments for considering MHD waves in this context, besides their ubiquitous nature.
Theoretical studies of time-dependent magnetic reconnection under coronal conditions show that Poynting flux 
is the dominant downwards energy flux beyond the diffusion region and reconnection jets \citep{2007LongcopePriest,2009Birn,2010Kigure}.
\citet{2009Birn} also demonstrated that Poynting flux can be particularly intense in tightly focused spots that
resemble the brightest chromospheric emission sources, the spots being associated with fragmentation of a 3D coronal current sheet.
Furthermore, the energy flux carried by MHD waves in active regions can be comparable to the flare energy flux for reasonable
perturbations because Alfv\'en speeds can be large (up to a few times $0.1c$) according to coronal field strengths 
deduced from gyrosynchrotron emission \citep{2006BrosiusWhite} or obtained by magnetic field extrapolation \citep{2008Regnier}
(estimates of Alfv\'en speeds from observations were discussed in detail by \citet{2008FletcherHudson}).
Magnetic field strengths and densities vary throughout active regions, 
and for most active regions the figure of a few times $0.1c$ refers to places where the Alfv\'en speed is particularly high,
however, these are also the parts of an active region where flare energy transport by Poynting flux is likely to be of most interest. 

That the magnetic field can be strongly perturbed during a solar flare is not in question: 
there are many observations of changes to both the line-of-sight and horizontal magnetic field at the photosphere at the time of the 
flare impulsive phase \citep[e.g.][]{2005SudolHarvey,2010WangLiu,2011Su,2012Wang}.
A broader question is what effect such perturbations, which are presumed to be driven by magnetic rearrangement in the corona, 
could have on the flare atmosphere.

MHD waves have long been contemplated to play a role in the energy balance of the chromosphere,
with \citet{1956Piddington} realizing that ion-neutral collisions in the chromosphere
produce a wave damping that is many orders of magnitude greater than that produced by either 
viscosity or classical resistivity.
This was first used to explain heating of the chromosphere by upgoing waves produced in the upper convection zone and photosphere
\citep{1961Osterbrock}, but later applied to solar flares by \citet{1982EmslieSturrock},
who studied how Alfv\'en waves might contribute to the heating of 
the temperature minimum region (TMR) that is inferred from optical and UV flare observations 
\citep{1978Machado,1979Cook,1979CookBrueckner,1990Metcalf_a}.  
Based on their single-fluid MHD, WKB treatment, \cite{1982EmslieSturrock} concluded that Alfv\'en wave damping is an energetically 
viable mechanism for TMR heating. 

The origin of temperature minimum heating and optical flare emission is still not fully explained. 
\cite{1978Machado} and \cite{1990Metcalf_b} provide excellent summaries of the candidate mechanisms, 
and \cite{1990Metcalf_b} then rules most of them out based on observations of 5 flares. 
Direct excitation of the TMR by electron beams, as proposed by  \cite{1986Aboudarham} and \cite{1987Aboudarham}, 
or by proton beams, is ruled out on the basis of inconsistency with the observed hard X-ray and $\gamma$-ray signatures. 
TMR heating by photoelectrons generated by EUV or soft X-ray photoionization of the TMR by an 
overlying coronal or transition-region source is also ruled out. 
The most likely mechanism is identified by \cite{1990Metcalf_b} as chromospheric backwarming, 
in which enhanced chromospheric continuum radiation produced by the flare is directly absorbed (by the H$^-$ ion) in the TMR, 
heating it and producing enhanced optical emission from the TMR and from the photosphere below (itself heated by the increased TMR emission). 
This mechanism is supported also by numerical simulations based on energy deposition and excitation by an electron beam from the corona
\citep[e.g.][]{2010Cheng} and by observational comparisons with calculations involving electrons accelerated more locally 
to the radiation site \citep{1999Ding}.
However the role played by Alfv\'en waves in chromospheric heating has not been explored to a comparable degree, 
so as part of this study we revisit the question of TMR heating by this mechanism.

It should be noted that the wave dissipation model for TMR heating was dismissed by \cite{1988Cheng} 
on the basis that the time delay predicted by \cite{1982EmslieSturrock} between the appearance of flare hard X-rays and the TMR heating, 
due to the wave travel time from corona or upper chromosphere to the TMR, was too long compared to observations. 
On the other hand,  \cite{1990Metcalf_b} ruled out the model on the basis that the predicted delay was a factor 2 too short 
to explain observations. 
The time delays depend  on the  local Alfv\'en speed in the chromosphere, 
which may be expected to vary strongly from place to place in an active region, 
and is dependent on the vertical variation of the chromospheric density and field strength, 
neither particularly well constrained by observations or models.

Typically, TMR temperature increases of between $100$ and $300$ K compared to typical quiet Sun
values have been reported \citep{1975MachadoLinsky,1990Metcalf_b}, 
located at a few hundred kilometers above the photosphere. 
The total energy flux required to power the optical emission from a large flare is of the order of $10^{8}\rm{~J~m^{-2}~s^{-1}}$ 
(deduced by dividing the radiated power in the flare reported by \cite{2010Watanabe} 
by an estimate of the flare's optical footpoint area from their Figure 2, 
while from \cite{2007Fletcher} we can set more modest (lower) limits of around $10^{7}\rm{~J~m^{-2}~s^{-1}}$, 
based on the measured UV footpoint area (which is, however, typically larger than the optical area).  
These are therefore the typical values that must be provided by the wave Poynting flux entering and dissipated in the chromosphere.

With solar flares as motivation, this paper examines Alfv\'enic wave propagation from the fully ionized corona
into and through the underlying partially ionized chromosphere.
Particular attention is given to the transmission of energy from the corona to the underlying atmosphere
and to damping of waves in the chromosphere through ion-neutral friction and neutral-enhanced electron resistivity.
These are crucial steps towards detailed understanding of the role MHD waves play in flares.

Recently, the propagation, damping and energy flux of Alfv\'enic waves in partially ionized plasma has been examined in a solar context by 
\citet{2003KumarRoberts, 2004Khodachenko, 2006Khodachenko, 2005Leake, 2008Pandey, 2008PandeyWardle, 2008Vranjes, 2011Zaqarashvili},
confirming earlier results that ion-neutral collisions are responsible for significant dissipation in the chromosphere.
Effects of partial ionization have also been considered for 
damping of waves in prominences \citep{2007Forteza, 2009Soler, 2010Soler, 2010Carbonell},
formation of spicules \citep{1992Haerendel, 1998DePontieuHaerendel, 2002JamesErdelyi, 2003James},
chromospheric heating by dissipation of magnetic flux tubes \citep{2012KhomenkoCollados},
and generation of electric currents by relative flows between ions and neutrals \citep{2010Krasnoselskikh}.
Additionally, the presence of neutrals may be important for 
magnetic flux emergence \citep{2006LeakeArber, 2007Arber}
and magnetic reconnection \citep{1989Zweibel, 2009SakaiSmith, 2011Zweibel, 2012Leake}.

Transmission of Alfv\'enic waves from the corona to the chromosphere and their subsequent damping
have previously been investigated analytically by \citet{2001DePontieu}, 
who evaluated damping times at different heights in model chromospheres and
computed energy transmission coefficients for a continuous Alfv\'en speed profile that is uniform in the photosphere and corona,
and exponentially growing with height in the chromosphere, using a constant scale height.
However, they did not include a step in Alfv\'en speed at the transition region, 
and therefore overestimated transmission from the corona to the chromosphere.

We revisit the subject using a two-fluid model (plasma and neutrals)
that avoids adoption of the WKB approximation or assumptions of Cowling conductance.
We also go beyond previous investigations by solving the complete time evolution of Alfv\'enic wave pulses launched from the corona,
using realistic density and temperature structure from a selection of semi-empirical atmospheric models by \citet{2009Fontenla}.
This approach allows the most detailed and realistic investigation of the problem to date.
Although this paper focuses on downgoing waves, 
the models and methods described can equally be used in understanding upgoing waves
and may therefore be valuable to study of the solar wind, coronal and chromospheric heating, and spicules.

\section{Energy Transmission at the Transition Region}\label{sec:tr_trans}

Before beginning detailed investigation, it is useful to estimate energy transmission across the solar transition region
for a simplified model where an Alfv\'en wave encounters a step in Alfv\'en speed, 
corresponding to the transition between a uniform corona and uniform chromosphere.
Although highly simplified, this model produces some useful insights.

In the corona, the wave speed is 
\begin{eqnarray}
 c_{cor}&=&\frac{B_0}{\sqrt{\mu_0 \rho_{cor}}},
\end{eqnarray}
where $B_0$ is the background magnetic field strength and $\rho_{cor}$ is the mass density of the coronal plasma.
For the chromosphere, we assume that neutral particles and ions are perfectly coupled by collisions, making the wave speed
\begin{eqnarray}
 c_{chr}&=&\frac{B_0}{\sqrt{\mu_0 \rho_{chr}}},
 \label{eq:cA_chrom}
\end{eqnarray}
where $\rho_{chr}$ is the total mass density (ions and neutrals).

Solving the linear momentum and ideal induction equations, 
subject to requirements that an infinitesimal region around the transition region has finite acceleration 
and is neither a source nor sink of wave energy, the energy transmission coefficient (fraction of incident wave energy that is transmitted) is obtained as
\begin{eqnarray}
 \mathcal{T} &=& \frac{4c_{cor}c_{chr}}{\left(c_{cor}+c_{chr}\right)^2}=\frac{4\phi^{1/2}}{\left(1+\phi\right)^2}.\label{eq:trans:T}
\end{eqnarray} 
Here, we have followed the example of \citet{1982EmslieSturrock} in introducing a dimensionless parameter, $\phi$, defined as
\begin{eqnarray}
 \phi &=& \left(\frac{c_{chr}}{c_{cor}}\right)^2=\frac{\rho_{cor}}{\rho_{chr}}=\frac{n_{cor}}{n_a+n_e},
\end{eqnarray}
where the last equality holds for hydrogen-dominated plasma with $n_{cor}$ the coronal plasma number density,
$n_e$ the chromospheric plasma number density, and $n_a$ the chromospheric neutral number density.
These results hold for waves incident from either the corona or chromosphere.

It is convenient to express energy transmission in terms of the chromospheric and coronal temperatures.
Assuming pressure balance at the transition region,
\begin{eqnarray}
 2n_{cor}T_{cor}=(n_a+2n_e)T_{chr}.
\end{eqnarray}
Hence,
\begin{eqnarray}
 \phi &=& \left(\frac{T_{chr}}{T_{cor}}\right)f\left(n_e/n_a\right)
\end{eqnarray}
where $f(n_e/n_a)=(0.5+n_e/n_a)/(1+n_e/n_a)$ is a monotonically increasing function,
bounded between $0.5$ (neutral chromosphere; $n_e/n_a\rightarrow 0$) and $1$ (fully ionized chromosphere; $n_e/n_a\rightarrow \infty$).
The parameter $\phi$ is typically of order $0.01$ and can therefore be neglected on the denominator of (\ref{eq:trans:T}).
Thus, the energy transmission coefficient, $\mathcal{T}$, lies in the range:
\begin{eqnarray}
 2\sqrt{2}\sqrt{\frac{T_{chr}}{T_{cor}}}<&\mathcal{T}&<4\sqrt{\frac{T_{chr}}{T_{cor}}},\label{eq:trans:bounded}
\end{eqnarray}
with $\mathcal{T}$ approaching the upper limit if the upper chromosphere is fully ionized,
and the lower limit if the upper chromosphere is dominated by neutrals.

The limits on $\mathcal{T}$ are easily evaluated for typical solar conditions.
For example, substituting $T_{chr}=10000\mbox{ K}$ and $T_{cor}=2\mbox{ MK}$ into (\ref{eq:trans:bounded}) 
gives the range $0.20<\mathcal{T}<0.28$.
We therefore expect significant transmission of wave energy across the transition region.
This is especially true if coronal waves are trapped (on closed field lines, within density structures or inside the chromosphere)
and therefore incident on the transition region multiple times.

The exact result is dependent on the temperatures chosen, however the dependence is lessened by the square root:
even if the temperature ratio were increased fourfold, 
the energy transmission coefficient would still be between $0.10$ and $0.14$.
If collisional coupling between chromospheric ions and neutrals were imperfect (which may be the case for high frequency waves)
the primary effect would be to increase $c_{chr}$ through reduced mass loading and hence increase transmission.
We also note that, although the present analysis is founded on Alfv\'en waves, 
it seems likely that similar conclusions will apply to other MHD waves (for example, fast or kink waves), 
something that should be confirmed by future work.

We conclude from this short analysis that waves with an Alfv\'enic nature can carry significant energy across the transition region.
This motivates further study, so we proceed to detailed modeling of Alfv\'en waves in a solar atmosphere with realistic height structure.

\section{Modeling Alfv\'en Waves in the Chromosphere}
\subsection{Chromospheric Properties}\label{sec:model:properties}

Semiempirical models for chromospheric composition, densities and temperature 
include the well-known models of \citet{1981VAL} and, more recently, \citet{2009Fontenla}.
The latter have state-of-the-art physics and distinguish between different 
solar features such as umbra, penumbra, plage and faculae.
We have chosen non-flaring models because we are interested in how the chromosphere might get into the flaring state from a preflare condition.

\begin{figure}
 \epsscale{0.6}
 \plotone{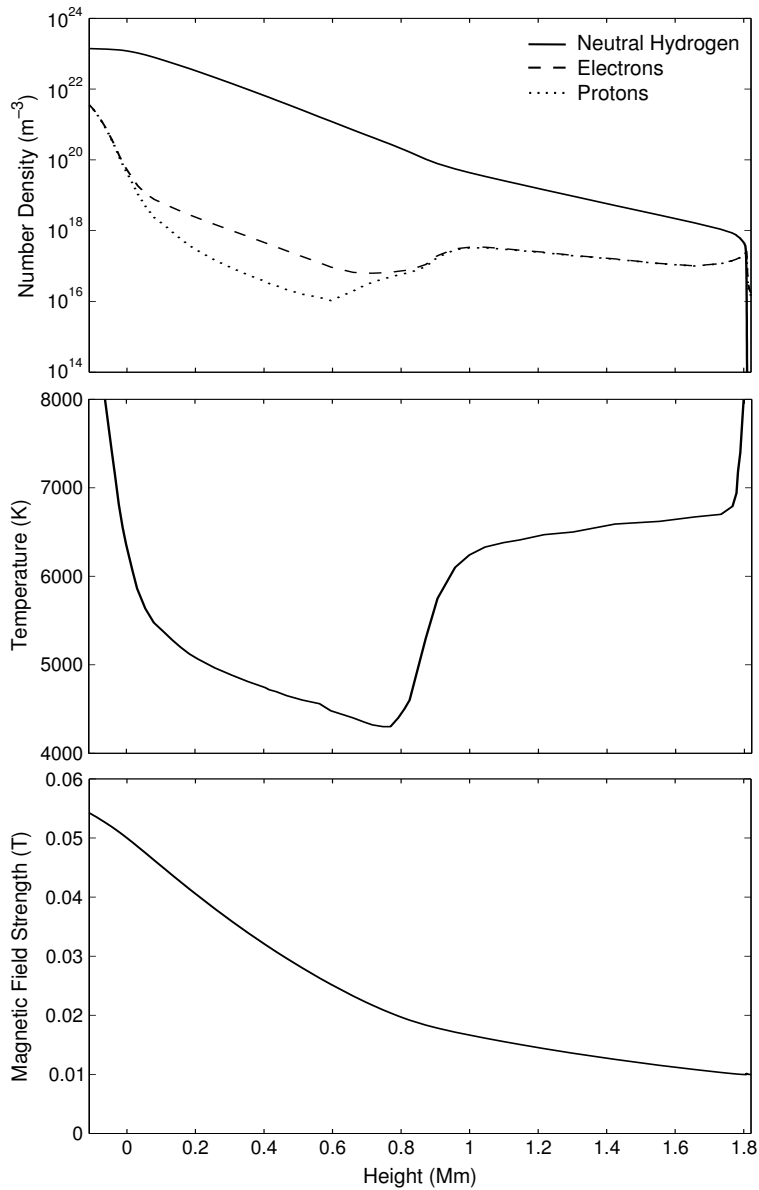}
 \caption{Plage chromospheric model.  
         (top) Number densities of neutral hydrogen, electrons and protons, from \citet{2009Fontenla}.  
         Where proton number densities depart from electron number densities, this is due to the presence of heavy ions.
         (middle) Temperature, from \citet{2009Fontenla}.
         (bottom) Magnetic field strength in a representative flux tube, assuming power law scaling with gas pressure.
\label{fig:plage_properties}}
\end{figure}

Figure \ref{fig:plage_properties} (top) and (middle) show particle number densities and temperature 
as functions of height for the plage model described by \citet{2009Fontenla}.
Neutral hydrogen is the most common particle and it is essentially in thermal hydrostatic equilibrium,
although the model does consider parametrically-described non-gravitational forces.
The neutral scale height is typically between 120~km (lower chromosphere) and 230~km (upper chromosphere). 
Electrons and ions are also present, 
with the fraction of hydrogen that is ionized varying from 
$5\times10^{-6}$ at the TMR to $1$ at the transition region.
At the TMR, Figure \ref{fig:plage_properties} (top) shows that
electron number density is an order of magnitude greater than proton number density.
The difference is made up by `heavy' ions that substantially 
outnumber protons at these heights because hydrogen ionization is so low.
Thus, magnesium, silicon and iron  ions -- the most abundant elements in the Sun with first ionization potentials lower than hydrogen's -- 
should be included in chromospheric models that include the TMR.

Magnetic field strength as a function of height in the chromosphere is not well established, 
due to the difficulty of making direct chromospheric measurements,
but it is widely agreed that magnetic field strength in flux tubes should decrease with height.  
In the absence of observationally-derived models, we follow an established precedent and model the decay in magnetic field strength as
\begin{eqnarray}
  B(z) &=& B(0)\left(\frac{P(z)}{P(0)}\right)^\alpha,
  \label{eq:B}
\end{eqnarray}
where $P(z) = k T (n_a+2n_e)$ is the total gas pressure \citep{1983ZweibelHaber}.
For a plage model, we take $B=0.05\mbox{ T}$ at the photosphere, in keeping with magnetogram measurements,
and $B = 0.01\mbox{ T}$ at the top of the chromosphere.
Combined with gas pressures from \citet{2009Fontenla},
this gives the magnetic field profile shown in Figure \ref{fig:plage_properties} (bottom), for which $\alpha=0.139$.

\subsection{Coronal Properties}\label{sec:model:corona}

Coronal properties partially determine wave transmission across the transition region,
and including a coronal domain in our model also allows propagation of an incident wave to and from the transition region.  
Neither of these roles require realistic height dependence away from the transition region,
hence we will model the corona as a uniform fully ionized plasma with properties appropriate to the low corona.
Throughout this work, a coronal temperature of 1~MK is assumed, and we impose a condition of gas pressure balance at the transition region,
from which coronal plasma density is deduced.  
The resulting coronal density varies between solar features from a few times $10^{15}\mbox{ m}^{-3}$
to $10^{16}\mbox{ m}^{-3}$ (over sunspots), in reasonable agreement with accepted values for the low corona.

We have also run additional simulations with lower coronal densities (assuming coronal temperatures of 2~MK and 4~MK).
Halving coronal density reduces transmission to the chromosphere by a factor of $\sqrt{2}$, 
in line with equation (\ref{eq:trans:bounded}), and does not alter our damping results.

\subsection{Governing Equations}\label{sec:model:eqs}
Section \ref{sec:model:properties} shows that the chromosphere must be described as a partially ionized plasma
with neutral and plasma components.  One way to do this is using a multi-fluid approach, which we outline here.

The neutral component of the plasma may be modeled as a fluid with the following momentum equation:
\begin{eqnarray}
  \rho_n \frac{D_n \vec{u}_n}{Dt} &=& -\nabla p_n +\rho_n\vec{G}
  +\rho_n\nu_{n,i}\left(\vec{u}_{i}-\vec{u}_n\right)
  +\rho_n\nu_{n,e}\left(\vec{u}_{e}-\vec{u}_n\right),
  \label{eq:momentum_neutral}
\end{eqnarray}
where $D_s/Dt\equiv\partial/\partial t + \vec{u}_s\cdot\nabla$ is the convective derivative.
Terms on the right-hand side represent, from left to right, pressure gradients, gravity
and momentum exchange with charged particles via collisions.
Here $\nu_{s,t}$ are momentum transfer collision frequencies, 
generally a function of temperature and the number density of target particles, $n_t$.
These have the property
\begin{eqnarray}
  \rho_s \nu_{s,t} &=& \rho_t \nu_{t,s}.
  \label{eq:frequency_swap}
\end{eqnarray}

Momentum transfer takes the form of a friction, as shown in equation (\ref{eq:momentum_neutral}),
provided relative velocities between particles are less than their average thermal velocities.
More generally, some pairs of species require $\nu_{s,t}$ to be multiplied by a correcting function of relative velocity, 
which reduces to unity in the linear limit we have taken \citep{2009SchunkNagy}.

We also work with the total momentum equation for the partially-ionized plasma as a whole:
\begin{eqnarray}
  \rho \frac{D \vec{u}}{Dt} &=& \vec{j}\times\vec{B} -\nabla p +\rho\vec{G};
\end{eqnarray}
low-frequency Amp\`ere's law:
\begin{eqnarray}
  \vec{j} &=& \frac{1}{\mu_0}\vec{\nabla}\times\vec{B};
  \label{eq:amperes}
\end{eqnarray}
Faraday's law:
\begin{eqnarray}
 \frac{\partial \vec{B}}{\partial t} &=& -\vec{\nabla}\times\vec{E};
  \label{eq:faradays}
\end{eqnarray}
and generalized Ohm's law:
\begin{eqnarray}
  \vec{E} &=& -\vec{u}_i\times\vec{B} + \frac{1}{\sigma_e}\vec{j} 
              + \frac{1}{ne}\left(\vec{j}\times\vec{B}-\vec{\nabla}p_e\right),
  \label{eq:ohms_gen}
\end{eqnarray}
with
\begin{eqnarray}
  \sigma_e &=& \frac{n_e e^2}{m_e\left(\nu_{e^-,i^+}+\nu_{e^-,H}\right)}
\end{eqnarray}
being the electron conductivity allowing for collisions with ions and neutrals 
(this is the same as the parallel conductivity in a single-fluid description of partially ionized plasma).
Finally, the single fluid velocity is related to the neutral and ion velocities by
\begin{eqnarray}
  \rho\vec{u} &=& \rho_n\vec{u}_n + \rho_i\vec{u}_i,
  \label{eq:stream_velocity}
\end{eqnarray}
where an electron contribution is neglected owing to small electron mass.

\subsection{One-Dimensional Linear Model}\label{sec:1d_model}
For the present investigation, the model is reduced to one dimension. 
1D models are useful tools that provide insight by addressing the key physics of a problem in a simplified way.
Their results, and lessons from their development, can also pave the way for 3D modeling
by establishing the most fundamental behavior of a system, 
proving techniques, and identifying potential problems before large resources are committed.
If properly constructed, their results are often robust (to at least leading order) when compared with higher dimensional models.

Our governing equations are simplified by assuming horizontal invariance, linearizing the governing equations and
neglecting the Hall term in Ohm's law (a valid approximation provided length scales 
remain longer than the ion skin depth, $\delta_i = \sqrt{m_i/(\mu_0 n_e e^2)}$, which is consistent with our results).
Under these conditions, horizontal velocity perturbations decouple from vertical motions 
and do not compress the plasma or magnetic field, therefore taking the form of shear Alfv\'en waves. 
One-dimensional modeling also requires that we take the equilibrium magnetic field direction as vertical,
so that there is a single preferred direction, 
and we neglect the effect of electron-neutral collisions in the neutral momentum equation
in favor of ion-neutral collisions, owing to the much smaller momentum carried by electrons than ions. 

The linear assumption limits our model to early times during the response of the solar atmosphere to incident coronal waves,
which is the focus of this study, and to wave amplitudes with $\delta B \ll B_0$.  
Relaxing this assumption is an important target for future work motivated by this study
and is discussed in Section \ref{sec:conclusions}.
Other goals for future work include introducing second and third dimensions and more sophisticated modeling of the 
interaction betweens neutral particles and the plasma.  These are also discussed in Section \ref{sec:conclusions}.

Using these simplifications, the equations governing shear Alfv\'en waves become
\begin{eqnarray}
  \frac{\partial u}{\partial t} &=& \frac{B_0}{\rho} j,\\
  \frac{\partial u_n}{\partial t} &=& \nu_{n,i}\left(u_{i}-u_n\right),\label{eq:dun}\\
  \frac{\partial j}{\partial t} &=& \frac{\partial^2}{\partial z^2}\left(\frac{B_0}{\mu_0}u_i+\frac{1}{\mu_0\sigma_e}j\right).
\end{eqnarray}
There are computational advantages to using the three variables $u$, $u_n$ and $j$ 
and these are described in Section \ref{sec:computing}.
The final step is to eliminate $u_i$, which is done with the aid of equation (\ref{eq:stream_velocity}).
Doing so, and making use of equation (\ref{eq:frequency_swap}), the governing equations can be expressed as
\begin{eqnarray}
  \frac{\partial u}{\partial t} &=& \frac{B_0}{\rho} j,\label{eq:solve:one}\\
  \frac{\partial u_n}{\partial t} &=& \left(\nu_{n,i}+\nu_{in}\right)(u-u_n),\label{eq:solve:two}\\
  \frac{\partial j}{\partial t} &=& \frac{\partial^2}{\partial z^2}\left(
          \frac{B_0}{\mu_0}\left(u+\frac{\rho_n}{\rho_i}\left(u-u_n\right)\right)
          +\frac{1}{\mu_0\sigma_e}j\right).\label{eq:solve:three}
\end{eqnarray}

For the chromosphere, equations (\ref{eq:solve:one}) to (\ref{eq:solve:three}) are solved as written; 
for the corona, equation (\ref{eq:solve:two}) is dropped 
and the limits for fully ionized plasma at large magnetic Reynolds number 
($\rho_n\rightarrow0$ and $1/\sigma_e\rightarrow0$) are applied to equations (\ref{eq:solve:one}) and (\ref{eq:solve:three})
before solving.

If the displacement current is retained in Amp\`ere's law for the corona,
then the right hand side of the ideal version of equation (\ref{eq:solve:three}) is divided by a factor of 
$1+c_A^2/c^2$ (not shown above).
Consequently, Alfv\'en waves travel at the relativistic Alfv\'en speed
$c_{A,rel}=c_A/\sqrt{1+c_A^2/c^2}$ and therefore cannot exceed the speed of light
\citep[e.g.][]{2008LysakSong,2010Russell}.
In more extreme cases than are considered here, 
a relativistic treatment of the wave propagation may be required, 
however, for the semiempirical models studied in this paper,
all Alfv\'en speeds are nonrelativistic and the displacement current can be safely neglected
(for reference, if $c_A=0.1c$, which is larger than the values we consider,
the relativistic correction changes the wave speed by less than 0.5\%).

The goal of this study is to estimate transmission of wave energy to the chromosphere and quantify chromospheric damping,
hence, the crucial aspects that must be treated realistically are wave speeds, electron resistivity and
damping due to ion-neutral collisions.
This leads to a conflict, because wave speeds and the effectiveness of ion-neutral collisions in wave damping both depend 
on the strength of the equilibrium magnetic field, which must be allowed to vary, even though we have neglected 
horizontal components when setting the equilibrium magnetic field direction to obtain equations (\ref{eq:solve:one}) to (\ref{eq:solve:three}).
To get speeds and damping right, we therefore allow $B_0$ to vary as a function of height when solving equations 
(\ref{eq:solve:one}) to (\ref{eq:solve:three}).
In consequence, the 1D model is not self-consistent in terms of equilibrium field strength and direction,
however we have been careful to evaluate the impact of this on our model, checking for artifacts,
and we believe it does not unduly affects our results.
For example, we have verified that: $\vec{\nabla}\cdot \vec{B}=0$ for perturbations;
energy and momentum are both conserved and transported by appropriate vectors;
and comparison to simulations using uniform field strength shows only the expected changes due to wave speed and damping rates.

\subsection{Collision Frequencies}

Chromospheric ions are modeled as a mixture of protons, magnesium, silicon and iron, with
the number density of heavy ions determined by quasi-neutrality and relative abundances taken from \citet{1995KayeLaby}.
Heavy ions are assumed to be once-ionized where their numbers are significant (around the TMR).
At chromospheric temperatures, the dominant interaction between neutral hydrogen and protons is resonant charge exchange,
which transfers momentum with an effective collision frequency
\begin{eqnarray}
 \nu_{H,p^+} &=& 2.65\times10^{-16}T\left(1-0.083\log_{10}{T}\right)^2 n_p.\label{eq:nu:Hp}
\end{eqnarray}
Neutral hydrogen and heavy ions interact through non-resonant, Maxwell molecule collisions, giving
\begin{eqnarray} 
  \nu_{H,M^+} &=& 2.11\times10^{-15}\left(1+1/A_{M}\right)^{-1/2} n_{M^+},\label{eq:nu:HM}
\end{eqnarray}
where $A_M$ is the atomic mass number of the ion.
Combining these, the total neutral-ion collision frequency is
\begin{eqnarray}
  \nu_{H,i^+}&=&\nu_{H,p^+}+\nu_{H,Mg^+}+\nu_{H,Si^+}+\nu_{H,Fe^+}.
\end{eqnarray}
Note that the form of equation (\ref{eq:nu:HM}) means $\nu_{H,i^+}$
is insensitive to changes in the composition of the heavy-ion fluid.
The above frequencies are given by \citet{2009SchunkNagy} (SN) with close agreement to \citet{Geiss1986} (GB).

A number of formulas for $\nu_{e^-,H}$ appear in the literature.
Here, SN quote 
\begin{eqnarray}
  \nu_{e^-,H}&=&4.5\times10^{-15}T^{1/2}\left(1-1.35\times10^{-4}T\right)n_H,\label{eq:nu:eH}
\end{eqnarray}
GB quote $\nu_{e^-,H} = 6.97\times10^{-14}T^{0.1}n_H$ and
\citet[][(P)]{1982Priest} quotes $\nu_{e^-,H} = 1.95\times10^{-16}T^{1/2}n_H$.
All three formulas give $\nu_{e^-,H}>\nu_{e^-,i^+}$ around the temperature minimum,
meaning that electron-neutral collisions dominate electron resistivity at these heights,
whereas electron-ion collisions dominate in the upper chromosphere and solar interior.
In the \citet{2009Fontenla} plage model, at heights where electron-neutral collisions are significant,
the value for $\nu_{e^-,H}$ from GB is greater than the value from SN by a factor slightly less than 2, 
while the value obtained from P is less than the value from SN by a factor that can be as large as ten.
We chose to use the SN formula because it produces the intermediate value where electron-neutral collisions are significant
and is in reasonable agreement with the GB value.
The standing uncertainty in the electron resistivity of the chromosphere's most neutral layers is noted,
but fortunately has only a small impact on the conclusions of this work.


\section{Simulation Methods}\label{sec:computing}
Equations (\ref{eq:solve:one}) to (\ref{eq:solve:three}) were solved by finite difference on a non-staggered grid,
with the selection of primary variables ($u$, $u_n$, $j$) avoiding grid decoupling.
Chromospheric length scales are much shorter than coronal length scales (due to different Alfv\'en speeds) so
two different grid spacings were employed.
Within each domain, $\partial^2/\partial z^2$ was evaluated as a three-point centered finite difference.
At the interface point, a five-point finite difference was used to maintain second order accuracy.

The ion-neutral coupling time scale, $1/(\nu_{n,i}+\nu_{i,n})$, 
is on the order of milliseconds for the chromosphere and we will investigate waves with periods of order seconds,
which makes equation (\ref{eq:solve:two}) stiff.
This is a significant consideration for selecting a numerical scheme 
and we have been successful using the ode15s MATLAB solver algorithm \citep{1997Shampine}.
Accuracy is controlled by automatic adjustment of the time step to satisfy 
a relative tolerance of $10^{-8}$, an absolute tolerance on velocities of $10^{-10}\mbox{ ms}^{-1}$ 
and absolute tolerance on current density of $10^{-10}\mbox{ Am}^{-2}$ \citep[e.g.][]{NumericalRecipes}.

For this combination of finite differences and numerical scheme, numerical dissipation and dispersion
become noticeable when about 30 points or less resolve one wavelength,
the primary signature being an overshoot behind a wavepacket.
Simulations presented in this paper have been checked for these artifacts,
and comparison against simulations using fewer grid points shows them to be well converged.
The two-grid approach was validated using simulations of Alfv\'en waves in a fully ionized plasma,
first in a uniform medium and then with a discontinuity in Alfv\'en speed at the grid interface.
In these tests, the change in grid spacing did not produce numerical reflections, 
provided the wave was spatially resolved in both domains.

Simulations are driven by launching a downgoing wave in the corona,
using a time-dependent boundary condition for $j$ that allows upgoing waves to escape without reflection.
Using the ideal limit of equation (\ref{eq:solve:three}) in the corona ($\rho_n\rightarrow0$ and $1/\sigma_e\rightarrow0$) 
and splitting it into the sum of a contribution from the upgoing wave and specified downgoing wave:
\begin{eqnarray}
  \frac{\partial j}{\partial t} &=& \frac{\partial j_\downarrow}{\partial t}
          +\frac{B_0}{\mu_0}\frac{\partial}{\partial z}\left(\frac{\partial u_\uparrow}{\partial z}\right).
  \label{eq:boundary_dj_split}
\end{eqnarray} 
For ideal Alfv\'en waves in a fully ionized corona, propagating in one direction, magnetic and velocity perturbations are related by
$b/B_0=\pm u/v_A$, where the plus sign is for downgoing waves and the minus sign is for upgoing waves.
Thus, $u_{\uparrow}$ can be written in terms of total fields $u$ and $b$, giving
\begin{eqnarray}
  \frac{\partial j}{\partial t} &=& \frac{\partial j_\downarrow}{\partial t}
          +\frac{1}{2}\left[\frac{\partial^2}{\partial z^2}\left(\frac{B_0 u}{\mu_0}\right)-v_A\frac{\partial j}{\partial z}\right].
  \label{eq:boundary_dj}
\end{eqnarray} 
The spatial derivatives in equation (\ref{eq:boundary_dj}) apply to the upgoing wave and
can therefore be evaluated as one-sided finite differences using points inside the domain, 
without causing numerical instability.  We have used second order expressions 
and find that these produce a very clean outgoing boundary condition.
 
The incident wave is specified through $\partial j_\downarrow/\partial t$ in equation (\ref{eq:boundary_dj}).
For the simulations presented in this paper, $j_\downarrow$ is a single wavepacket that is both smooth and continuous:
\begin{eqnarray}
  \frac{\partial j_\downarrow}{\partial t}&=&\left\{
    \begin{array}{@{}rrcccl@{}}
       (2\pi j_0/\tau)\sin(4\pi t/\tau), \qquad &    0    &\leq& t &<& \tau/4\\
       (2\pi j_0/\tau)\cos(2\pi t/\tau), \qquad &  \tau/4 &\leq& t &<& 3\tau/4\\
       -(2\pi j_0/\tau)\sin(4\pi t/\tau),\qquad & 3\tau/4 &\leq& t &<& \tau\\
       0,                                \qquad &  \tau   &\leq& t &&
    \end{array}\right.
  \label{eq:driver}
\end{eqnarray}
where $\tau$ is the driver duration and $j_0$ is the current-density amplitude of the downgoing wave. 
The resulting $j(t)$ is initially zero, grows to $j=j_0$ at $t=\tau/4$, declines to change sign at $t=\tau/2$,
reaches $j=-j_0$ at $t=3\tau/4$ and returns to zero at $t=\tau$.

Simulations were performed on four cores of an i5-750 processor (clock speed of 2.66 GHz) with 4 GB of RAM.
Run times ranged from less than one hour up to three days,
the longer running simulations being those requiring the more grid points to resolve smaller length scales.

\section{Case Study: Plage Atmosphere, 1 Second Driver}\label{sec:case_study}
A case study is now presented, using a plage model and a 1 second duration driver.
This allows for detailed discussion of the evolution, the features of which are common to all drivers and atmospheric models.
Wave transmission to the chromosphere and damping within it are also quantified.
Having established key behaviors in this section, a parameter study will be presented in Section \ref{sec:parameters}
which examines the effects of wave period and atmospheric model.

We begin by describing the how the wave evolves in the case study.
First, the driver generates a downgoing Alfv\'en wave in the corona that partially reflects from the transition region.
The majority of wave energy (78.0\%) is reflected to form an upgoing Alfv\'en wave in the corona, which escapes from the simulation,
with the remaining 22.0\% transmitted to the chromosphere.

\begin{figure}
 \epsscale{0.8}
 \plotone{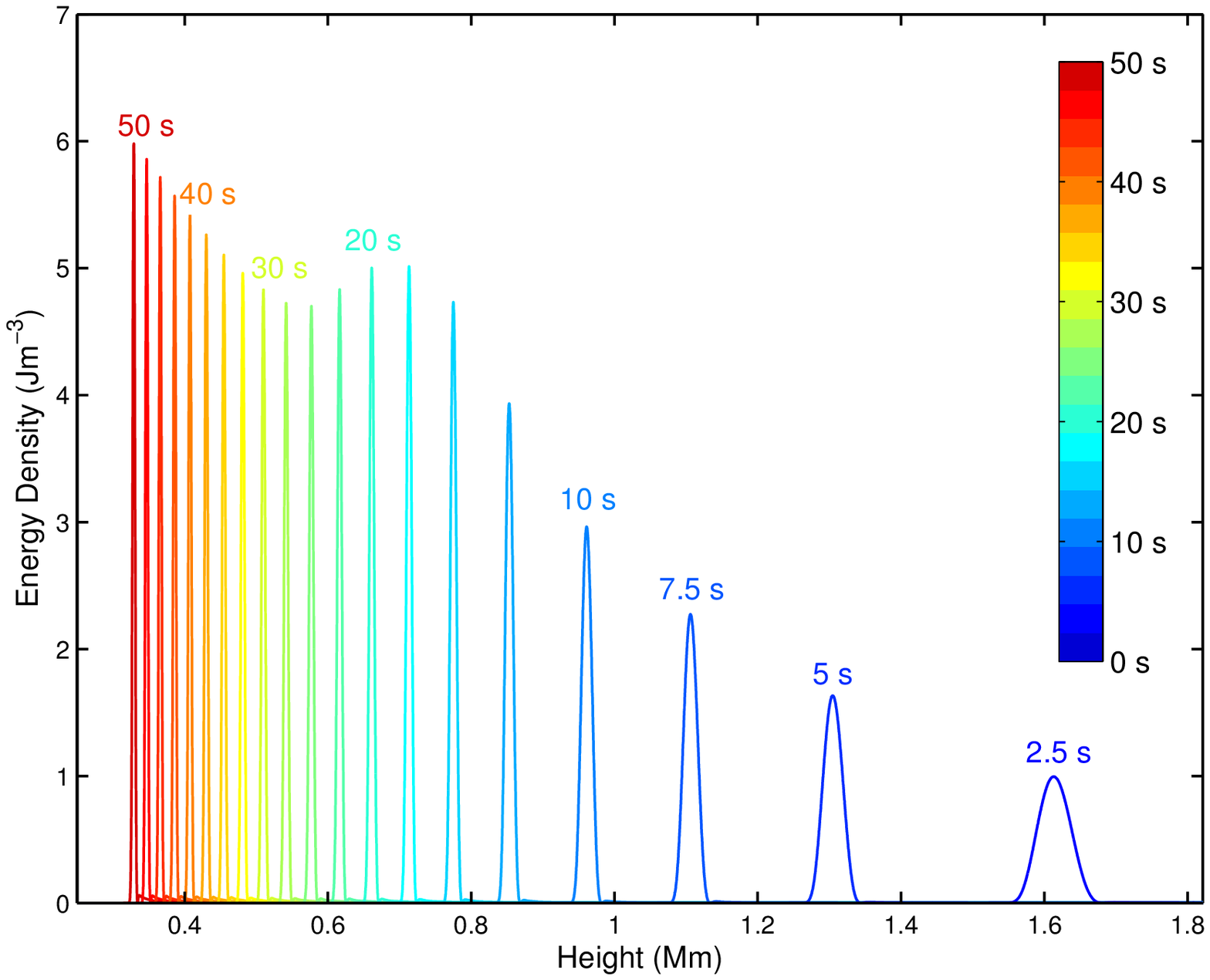}
 \caption{Snapshots showing evolution of energy density as a function of height 
  above the photosphere for case study of a 1 second period wave in the plage model.
  See the electronic edition of the Journal for a color version of this figure.
 \label{fig:snapshots}}
\end{figure}

Once in the chromosphere, the transmitted wave evolves as shown in Figure \ref{fig:snapshots},
which plots wave energy density (the sum of magnetic and kinetic energy densities) 
as a function of height at 2.5 second intervals. 
The wave propagates downwards through the chromosphere with very little additional reflection
because the width of the pulse (approximately 100~km at its widest; $t=2.5\rm{~s}$) is less
than the length scale of Alfv\'en speed gradients.
Here, $c_A\sim B n_H^{-1/2}$ changes primarily through $n_H$, which has a scale height of about 230~km in the upper chromosphere 
(see equation (\ref{eq:cA_chrom}) and Figure \ref{fig:plage_properties}),
thus the Alfv\'en speed length scale would be approximately 460~km if magnetic field were ignored.
In fact, magnetic field strength increases with decreasing altitude, 
partly compensating for the decrease in $c_A$ caused by increasing $n_H$, 
so the true length scale of Alfv\'en speed is longer than 460~km.
It follows that the spatial scales of the wave perturbations and equilibrium are well-separated in this case,
hence, the chromospheric wave evolves as in a WKB description and is not significantly reflected by Alfv\'en speed gradients.

Decreasing Alfv\'en speed does affect the Alfv\'en wave by slowing the pulse and altering its shape.
The trailing edge travels faster than the leading edge,
thus wavelength decreases with height, which in turn increases maximum wave energy density
(although height-integrated total energy density is conserved).

Of the wave fields, current density is most affected by the shortening wavelength. 
Magnetic-field perturbations increase as the wavelength shortens 
(to maintain height-integrated magnetic energy if damping rates are not large),
and current density is proportional to magnetic-field perturbation divided by wavelength.
Thus, current density would be inversely proportional to the square of wavelength in the absence of damping.
In this simulation, which includes physical damping, 
maximum current density at $t=50\rm{~s}$ is 28.9 times greater than maximum current density at $t=2.5\mbox{ s}$.

\begin{figure}
 \epsscale{0.8}
 \plotone{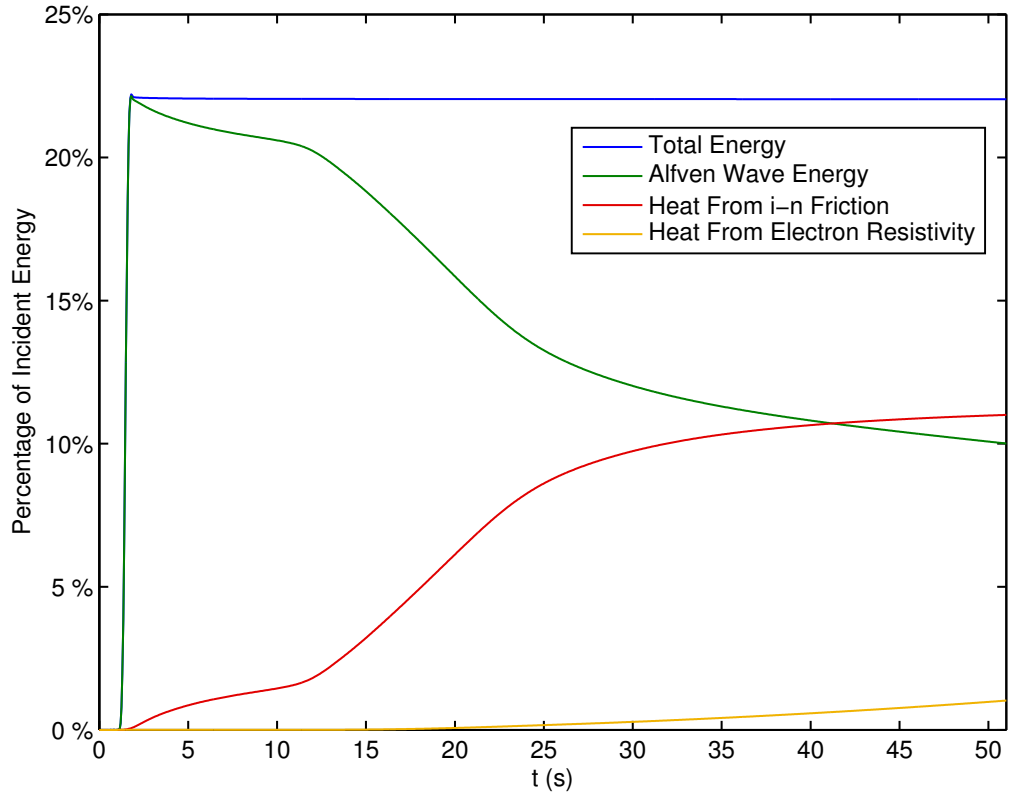}
 \caption{Energetics of the chromospheric domain for case study of a 1 second period wave in the plage model.
  The wave is launched in the corona at $t=0\mbox{ s}$ and enters the chromosphere at $t=1\mbox{ s}$.
  See the electronic edition of the Journal for a color version of this figure.
 \label{fig:energy_chromosphere}}
\end{figure}

Figure \ref{fig:energy_chromosphere} plots energy densities integrated in height over the chromospheric domain
and expressed as a fraction of the energy injected to the simulation by the incident coronal Alfv\'en wave.
One second into the simulation (equal to the travel time from the simulation's upper boundary to the transition region)
the energy in the chromosphere sharply increases as the wave enters the chromosphere.
In this case, 22.0\% of incident wave energy is transmitted to the chromosphere 
(evaluated using energy at the end of the simulation).
This is high enough to be of considerable interest.

As expected from analysis of Figure \ref{fig:snapshots},
Figure \ref{fig:energy_chromosphere} shows that very little energy is returned to the corona by subsequent reflections inside the chromosphere,
the total energy in the chromosphere remaining very nearly constant after the initial rise.

From the first moment when the Alfv\'en wave is in the chromosphere, it loses energy.
The decrease in wave energy is gentle when the wave is in the upper chromosphere, 
then rapid as it passes through the TMR, and more again gentle below this.
Energy lost from the Alfv\'en wave becomes heat, 
and 54.6\% of (non-equilibrium) energy in the chromosphere at the end of the simulation is heat that has been produced by wave damping.

To identify the source of this damping, we note that heating occurs at a rate 
\begin{eqnarray}
 \frac{\partial q}{\partial t} &=& m_Hn_H\nu_{n,i}(\vec{u}_n-\vec{u}_i)^2+\frac{\vec{j}^2}{\sigma_e}
 \label{eq:heating}
\end{eqnarray}
where the first term on the right-hand-side corresponds to ion-neutral friction 
and the second term is Ohmic heating due to electron resistivity.
Figure \ref{fig:energy_chromosphere} plots both heat contributions.
When the distinction is made, it is clear that wave damping is primarily due to ion-neutral friction, 
although electron resistivity plays a non-negligible role towards the end of the simulation.

\begin{figure}
 \epsscale{0.8}
 \plotone{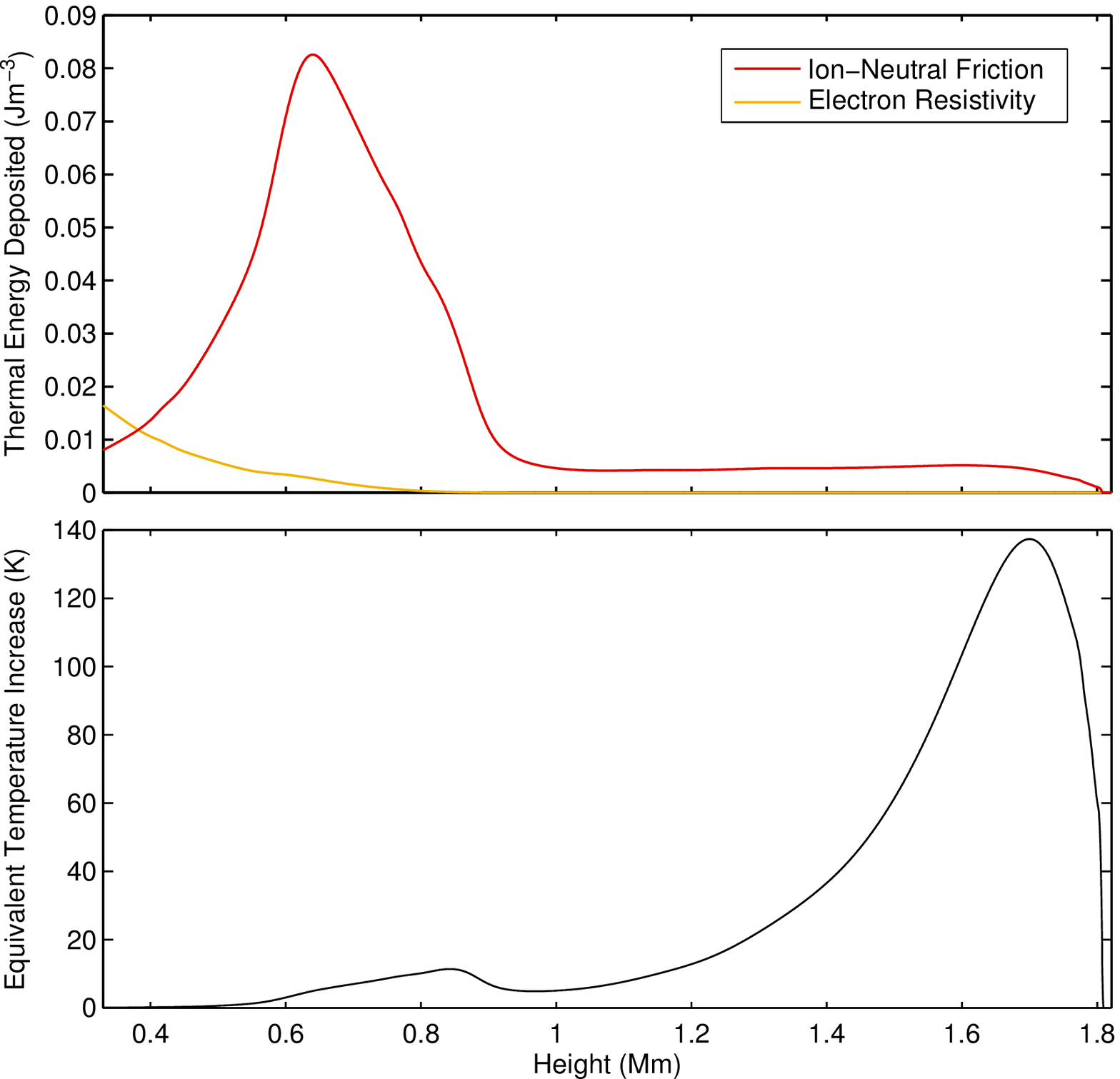}
 \caption{(top) Thermal energy deposited by a 1 second period wave in the plage model, through ion-neutral friction and electron resistivity.
          (bottom) Equivalent temperature increase produced by damping of the wave packet in the chromosphere 
          (thermal energy per particle).
          In (top) and (bottom), values correspond to a wave with coronal amplitude $\delta B/B_0=0.05$ and scale with $(\delta B/B_0)^2$.
          See the electronic edition of the Journal for a color version of this figure.
          \label{fig:heating}}
\end{figure}

Figure \ref{fig:heating} examines the locations of heating, with
Figure \ref{fig:heating} (top) showing the thermal energy produced by ion-neutral friction
and electron resistivity as a function of height.
It confirms that ion-neutral friction is the dominant heating mechanism
throughout most of the chromosphere, with electron resistivity taking over below about 0.4 Mm.
Very noticeably, the thermal energy deposited by wave damping peaks dramatically between about 0.4 to 0.9 Mm, 
which corresponds to the TMR.
Further investigation has shown that ion-neutral friction peaks in this layer
because it is where the ion-neutral coupling time is longest (i.e. $\nu_{H,i^+}$ is smallest).
This means that the relative velocity between ions and neutrals is greatest here,
and this increases the heating rate in equation (\ref{eq:heating}). 
(The increase in relative velocity enters equation (\ref{eq:heating}) as a square and therefore
more than compensates for the reduction in $\nu_{H,i^+}$.)
There is also a plateau of heating in the upper chromosphere, again due to ion-neutral friction.
Meanwhile, below the TMR, electron resistivity becomes significant, 
partly due to the many-fold increase in $j$ that we have described.  

Figure \ref{fig:heating} (bottom) shows the equivalent temperature increase 
that corresponds to the heat deposited by this wave (energy deposited per particle).
We do not attempt to address the thermodynamics and the loss terms that would ultimately govern the temperature profile,
however it is instructive to make a simple direct estimate
which serves as a crude indication of the likely locations and extent of temperature changes.
Although considerably more thermal energy is deposited at the TMR than in the upper chromosphere,
heating the upper chromosphere produces a larger change in temperature because of the much lower particle density.
Indeed, for a modest amplitude coronal wave with a magnetic perturbation of 5\% of the background field ($\delta B/B_0=0.05$),
a single 1 second pulse heats the upper chromosphere by a peak amount equivalent to a temperature increase of 137~K.
A secondary peak is also present close to the temperature minimum, 
where the heating is equivalent to 11.4~K in spite of the relatively high densities.
Values for other wave amplitudes can be obtained by scaling energy density and temperature 
as $(\delta B/B_0)^2$ since we are using a linear model.
These temperature increases are large enough to be interesting for chromospheric emission, 
discussed further in Section \ref{sec:conclusions}.

\section{Parameter Study}\label{sec:parameters}
Having established that transmission of wave energy from the corona to the chromosphere can be significant
and that chromospheric waves can be substantially damped (heating the chromosphere)
we now investigate sensitivity to different wave periods and atmospheric models.
 
Chromospheric models were based on the semiempirical models of \citet{2009Fontenla},
with magnetic field strength models provided by equation (\ref{eq:B}).
The six models are therefore determined by the parameters shown in Table \ref{tab:models}.
Simulations were performed for each of these models, with wave periods between 0.5 and 30~seconds.
In all cases, simulations were qualitatively similar to the case study
but differed in the fraction of energy transmitted to the chromosphere and 
the effectiveness of wave damping.

\begin{deluxetable}{ccll}
\tabletypesize{\scriptsize}
\tablecaption{Chromospheric models used for parameter study.}
\tablewidth{0pt}
\tablehead{
\colhead{Description} & \colhead{Fontenla Model ID} & \colhead{$B_{phot}$ (T)} & \colhead{$B_{cor}$ (T)}
}
\startdata
Umbra & 1006 & 0.2 & 0.05 \\
Weaker Umbra & 1006 & 0.2 & 0.025 \\
Penumbra & 1007 & 0.1 & 0.02 \\
Plage & 1004 & 0.05 & 0.01 \\
Facula & 1005 & 0.05 & 0.01 \\
Bright Facula & 1008 & 0.05 & 0.01
\enddata\label{tab:models}
\end{deluxetable}

\subsection{Energy Transmission From Corona to Chromosphere}
\begin{figure}
 \epsscale{0.9}
 \plotone{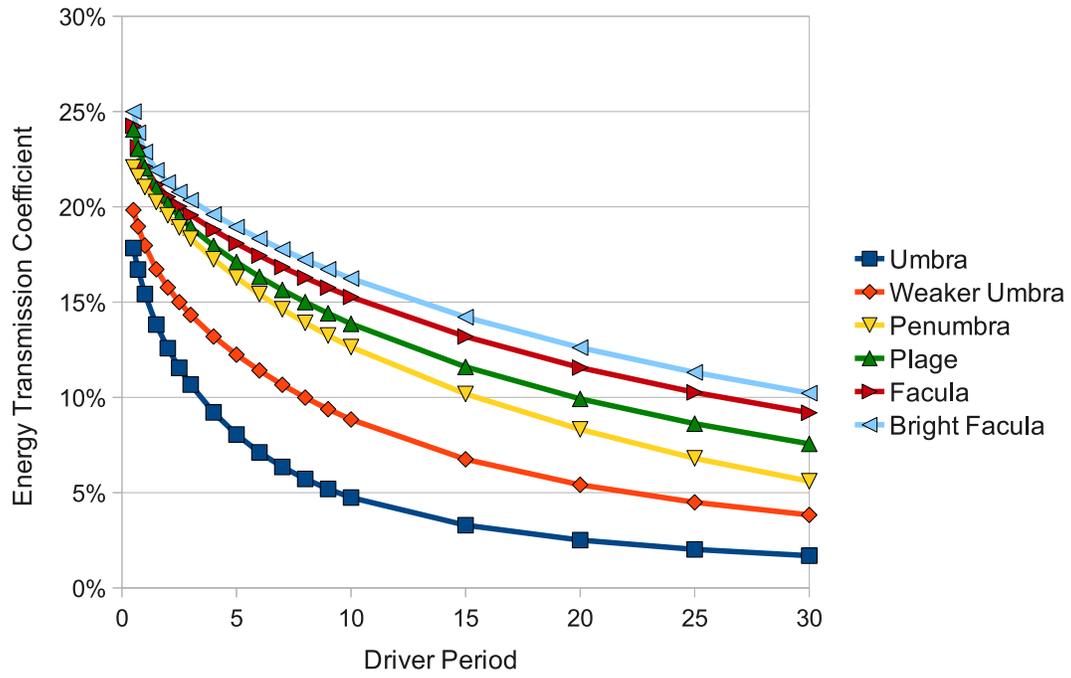}
 \caption{Energy transmission coefficients for waves incident from the corona.
          For each atmospheric model, transmission is shown for driver periods spanning 0.5 to 30 seconds.
          See the electronic edition of the Journal for a color version of this figure.
          \label{fig:compare_transmission}}
\end{figure}

Figure \ref{fig:compare_transmission} shows energy transmission to the chromosphere for each simulation,
evaluated as the energy that is located in the chromosphere at late times divided by 
the energy injected to the simulation by the incident coronal Alfv\'en wave.

The most striking feature is a strong dependence on wave period, 
transmission being greater for shorter period (higher frequency) waves.
Looking at the plage model, energy transmission is 24.1\% for a 0.5~s period wave, 22.0\% for a 1~s period,
13.9\% for a 10~s period and 7.6\% for a 30~s period.
This variation occurs because wave reflection is not determined solely by transmission across the transition region,
but also by wave properties in the chromosphere.

The transmission analysis of Section \ref{sec:tr_trans} is a good approximation 
when wavelengths are much shorter than coronal and chromospheric speed gradient length scales.
In this short-period limit, waves propagate without reflection once in the chromosphere,
as was seen in the 1 s period case study.
This places an upper limit on the energy transmission which, using equation (\ref{eq:trans:bounded}),
is 23.2\% for the plage model (1~MK corona and 6700~K neutral-dominated chromosphere).
In fact, the shortest period waves can beat this limit, which we attribute to the presence of a thin
layer of fully-ionized plasma at the base of the transition region, 
equation (\ref{eq:trans:bounded}) having shown that transmission to an 
ionized plasma is higher than transmission to a neutrally-dominated plasma. 

The assumption of uniform media breaks down when waves in the upper chromosphere have a wavelength comparable to or larger than the 
speed gradient length scale, in which case waves become subject to additional partial reflections that occur continuously.
In the plage model, 1 second period waves have a wavelength that is indeed much smaller than chromospheric scale heights,
so transmission approaches the limit described by equation (\ref{eq:trans:bounded}).
Longer period waves, however, are reflected in the upper chromosphere,
and the longest period waves are subject to the greatest additional reflection, as evident in Figure \ref{fig:compare_transmission}.

Since wave speed decreases with height (as a result of rapidly increasing density),
wavelengths decrease significantly as waves propagate through the chromosphere, as seen in Section \ref{sec:case_study}.
Scale heights do not change as significantly, 
e.g. staying in the range 120~km (lower chromosphere) to 230~km (upper chromosphere) in the plage model. 
Therefore, waves cease to reflect as they reach lower altitudes, hence
reflection within the chromosphere only happens in its upper reaches.

Upgoing waves produced by continuous reflection in the upper chromosphere partially reflect at the transition region to form new downgoing waves.
Therefore, wave energy is not easily lost from the chromosphere and multiple reflections can occur.
These partial reflections act to extend the downgoing wave by superposition behind the primary pulse.

Transmission also varies for different solar features.
The umbral models are notably different because they have a cooler chromosphere (minimum temperature of 3590~K) and are therefore denser.
They also have a lower transition region than other models, at about 1 Mm altitude, 
and do not have the upper temperature plateau found in other models (e.g. see Figure \ref{fig:plage_properties} (middle)).
This reduces transmission in two ways.

First, temperature contrast between corona and chromosphere -- and therefore density contrast -- 
is greater for umbral models than for models of other solar features.
The basic analysis of Section \ref{sec:tr_trans} suggests that energy transmission across the transition region scales as the square root 
of the temperature just below the transition region.
Given that temperatures below the transition region are about 4400~K in the umbral model and 6700~K in the plage model,
we therefore expect energy transmission for umbral models to be approximately 0.8 times that for the plage case
(for short period waves and at common coronal temperatures).  
This is in line with the reduction seen in simulations using a 1~s period driver,
from 22.0\% transmission for plage to 15.4\% transmission for umbra and 18.0\% transmission for the weaker umbra model.

The second consideration is that since the hydrostatic scale height is proportional to temperature, 
the cool umbral chromosphere has a much shorter density scale height (about 88~km) than do other solar features
(e.g. 230~km in the upper plage chromosphere).
Waves are therefore subject to greater reflection within the umbral chromosphere than within any other feature.
They are affected for longer durations, and reflection affects shorter period waves,
because a shorter wavelength is needed to turn off reflection.
It follows that the greatest differences in transmission to the chromosphere, when comparing different solar features,
occur at longer periods (e.g. for a 30 s period wave, transmission is 7.6\% in the plage model and a meager 1.7\% in the umbra model).

The differences between the umbra and weaker umbra models are due to the magnetic field model.
Both have a field strength of 0.2 T at 0 Mm,
but 0.025 T is applied at the top of the weaker umbra chromosphere, compared to 0.05 T at the top of the umbra chromosphere.
Thus, magnetic field strength changes more over the same distance 
and therefore has a shorter length scale in the weaker field model (the factor is 1.5).
Looking at equation (\ref{eq:cA_chrom}), increases in field strength with depth
counteract decreases in Alfv\'en speed caused by increasing density,
so Alfv\'en speed decreases more gradually in the weaker umbra model than in the umbra model.
Thus, in the weaker field model, waves experience less reflection within the chromosphere than for the stronger field model,
increasing overall transmission.

\subsection{Chromospheric Damping}

\begin{figure}
 \epsscale{0.9}
 \plotone{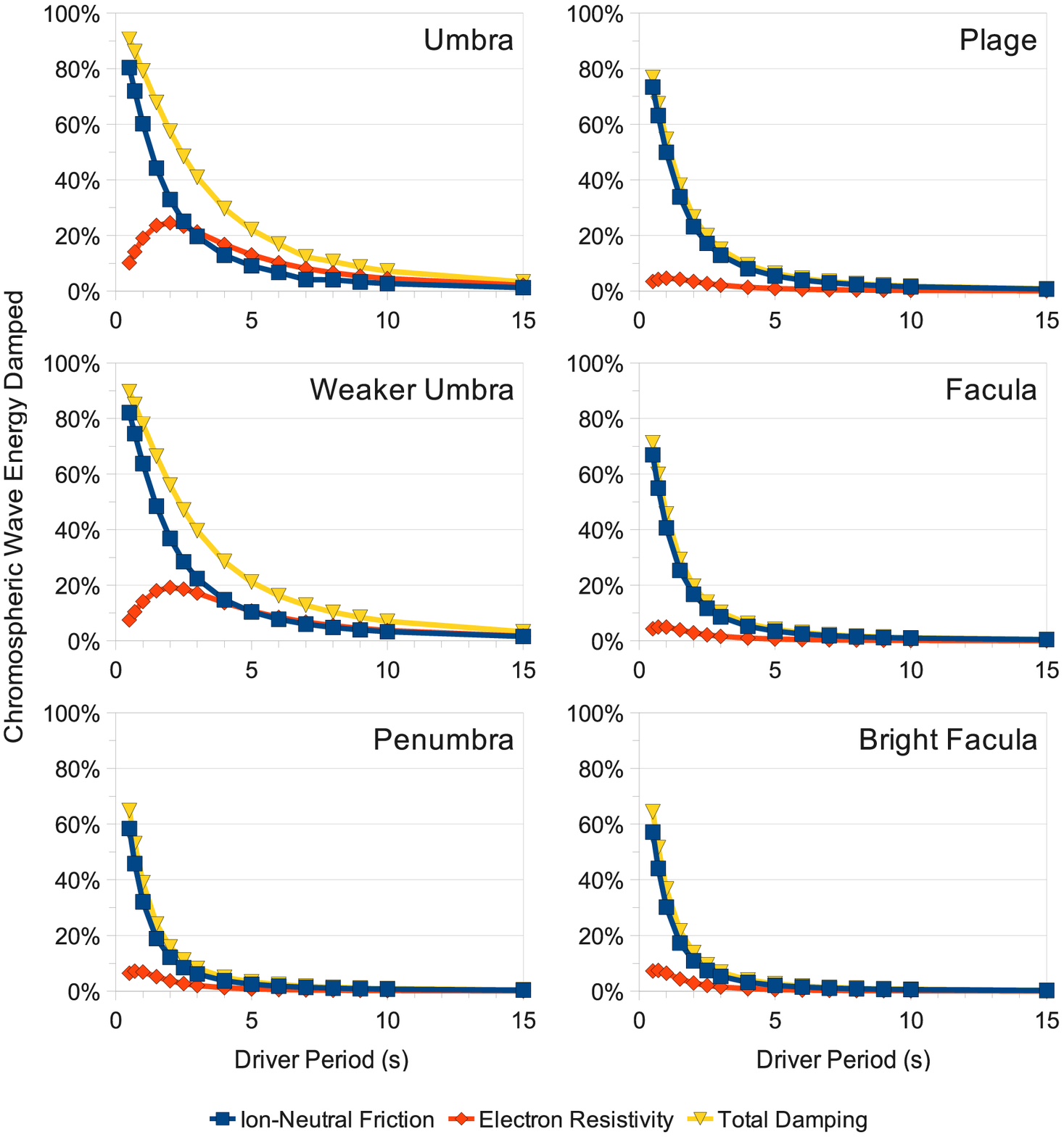}
 \caption{Percentage of chromospheric wave energy converted to heat by ion-neutral friction and (neutral enhanced) electron resistivity.
          For each atmospheric model, damping is shown for driver periods spanning 0.5 to 15 seconds.
          See the electronic edition of the Journal for a color version of this figure.
          \label{fig:compare_damping}}
\end{figure}

Chromospheric damping for different drivers and atmospheric models is shown in Figure \ref{fig:compare_damping},
which plots the percentage of energy in the chromosphere that is heat, at the end of each simulation.
Contributions from ion-neutral friction and electron resistivity are both identified. 

For 1~s period waves, between 36.6\% and 79.2\% of wave energy injected into the chromosphere has damped by the end of each simulation.
This damping is primarily due to ion-neutral friction, with the greatest losses occurring around the TMR,
although electron resistivity is responsible for up to a quarter of the total damping of 1~s waves in an umbra.

Damping is very sensitive to wave period.
In the plage model, 76.8\% of wave energy entering the plage chromosphere is damped for 0.5~s periods;
at 1~s periods damping falls to 54.6\%; 
and at 10~s periods, only 1.8\% is damped.
Thus, waves with periods of about 10 seconds or greater pass through the chromosphere to the interior relatively undamped,
while waves with periods of about 1 second or shorter are strongly damped and heat the chromosphere effectively.

A simple estimate of the damping time due to ion-neutral collisions illuminates the sensitivity of damping to wave period.
Using equations (\ref{eq:dun}) and (\ref{eq:heating}), the wave loses energy through ion-neutral friction at a rate
\begin{eqnarray}
 \frac{\partial E_{wave}}{\partial t} &=& -\rho_n\nu_{n,i}(u_n-u_i)^2 = \rho_n \nu_{n,i}\left(\frac{1}{\nu_{n,i}}\frac{\partial u_n}{\partial t}\right)^2.
 \label{eq:damping_estimate_one}
\end{eqnarray}
The derivative $\partial u_n/\partial t$ can be approximated as $u_n/\tau_{wave}$ where $\tau_{wave}$ is the wave period.
Also note that kinetic energy of neutrals makes up almost all kinetic energy in the chromosphere
and that kinetic and magnetic energies are approximately equal once waves are free to propagate downwards
(confirmed against our simulations), thus, $\rho_n u_n^2 \approx E_{wave}$, and hence,
\begin{eqnarray}
 \frac{\partial E_{wave}}{\partial t} &\approx& -\frac{E_{wave}}{\nu_{n,i}\tau_{wave}^2},
\end{eqnarray}
making the energy damping time
\begin{eqnarray}
  \tau_{damping}&\sim&\nu_{n,i}\tau_{wave}^2.\label{eq:damping_estimate}
\end{eqnarray}
Although this estimate is missing a coefficient of order unity (which will depend on the shape of the wave), 
we see that ion-neutral friction is most effective for short period waves and small values of $\nu_{n,i}$ (longer coupling times).
Of these two parameters, damping is more sensitive to wave period, which enters as a square.

Comparing different solar features across Figure \ref{fig:compare_damping}, damping is strongest in the umbral models, 
where almost 80\% of wave energy entering the chromosphere is converted to heat.
Ion-neutral damping is more effective in umbras than for other features because the neutral-ion coupling time is longer 
here -- typically one order of magnitude longer when compared with a plage model.
This allows larger relative velocities between ions and neutrals, which results in greater heating.
Damping due to electron resistivity is also substantially stronger in umbras than other solar features.
Comparing the umbra and plage models,
electron resistivity ($1/\sigma_e$) at 0.75 Mm altitude is similar in both models,
however it sharply diverges below this height, becoming as much as 250 times greater in the umbra model
and this causes greater damping of chromospheric waves (see equation (\ref{eq:heating})).

\section{Conclusions}\label{sec:conclusions}

Our studies have shown:
\begin{enumerate}
 \item Coronal Alfv\'enic waves can transmit a significant fraction of their energy to the chromosphere.
       Energy transmission is greatest for short period waves and can exceed 20\% 
       per incidence for waves with second or subsecond periods.  
       Cumulative transmission will be higher if coronal waves 
       are trapped in closed coronal structures (e.g. closed magnetic field),
       leading to multiple incidences.
 \item Waves in the chromosphere are subject to damping by ion-neutral friction and enhanced electron resistivity,
       with ion-neutral friction dominating in most cases.
       Damping is highly sensitive to period: waves with subsecond periods are damped effectively 
       (for 0.5~s periods up to 91\% of wave energy entering the chromosphere is converted to heat),
       whereas waves with periods longer than 10~s pass through the chromosphere with relatively little damping 
       (less than 7.2\%).
 \item Chromospheric damping of downgoing waves delivers most energy to the temperature minimum region (TMR), heating it directly.
       Energy is also deposited throughout the upper chromosphere, 
       where temperature increases are expected to be larger than at the TMR because number densities are much lower.
\end{enumerate}


These findings suggest that flare-generated MHD waves may be a viable means of transporting
flare energy from the corona to the chromosphere and heating the chromosphere.
They would have greatest effect if a substantial part of wave power were at second or subsecond periods.
Periods typically associated with solar flares, that one might expect to find in flare-generated waves,
range from subpulses in X-ray, radio and visible light observations (order of 100~ms), 
through elementary flare bursts (tens of seconds),
to the duration of the rise phase (several minutes).
There is, we conclude, reasonable overlap between observed flare time scales 
and the wave periods that have potential to heat the chromosphere.

For a case study of a single 1 second wave pulse in a plage model, 
a modest magnetic field perturbation of 5 G on a 100 G coronal background field
produced temperature increases of up to 137~K in the 
upper chromosphere and 11.4~K at the TMR.
Heating could be considerably amplified by increases in perturbation amplitude or background field strength.
If a train of such pulses were sustained for a minute,
the resulting temperature changes would certainly be significant.

It is interesting that the bulk of thermal energy is delivered to the TMR
(although temperature rises there are less than those in the less dense upper chromosphere).
This profile of heating is very different to that for electron beams, which lose their energy higher in the chromosphere.
Waves therefore offers a direct mechanism for heating the TMR, where the bulk of white light flare emission is formed,
that should be considered as a complement to backwarming.


Wave energy in the chromosphere that is not lost to heat is available to particle acceleration, since 
Alfv\'en waves with fields that vary perpendicular to the background magnetic field 
can carry magnetic-field-aligned electric fields 
created by inertial or kinetic effects \citep{2000Stasiewicz} or electron resistivity.
They could, therefore, potentially contribute to producing fluxes of energetic electrons in the 
tightly focused X-ray footpoints of a solar flare.
This may help to relieve theoretical difficulties associated with electron-beam-only collisional-thick-target models,
such as apparent requirements for large particle numbers, strong anisotropy and
high beam densities in the corona that would be unstable \citep{2008FletcherHudson,2009Brown,2011Krucker}.

Electron acceleration can only take place in the chromosphere if there is a population of electrons
for which energy losses to collisions can be overcome by energy gains between collisions.
This requires some electrons to have speeds significantly greater than typical quiet chromosphere thermal speeds 
(for which collisional losses are severe).
Two ways in which this can be achieved are by: 
(i) heating the chromosphere \citep{2008FletcherHudson}, e.g. by damping of flare-generated MHD waves; 
or, (ii) by having electrons present that previously precipitated from the corona
and retain sufficient velocity following each collision to be reaccelerated before the next collision \citep{2009Brown}.

In the case of directly accelerated chromospheric electrons, one can imagine a two-stage evolution, 
in which the upper chromosphere is initially heated by waves,
then energy is increasingly given over to acceleration of chromospheric electrons
as the high energy part of the electron distribution becomes fast enough for acceleration to overcome collisional losses.
A similar picture was presented in the flare burst model of \citet{2009LiuFletcher}.
This would be consistent with observations that some flares 
show increased thermal emission up to minutes before the rise phase of hard X-ray (HXR) emission.

In such scenarios, heating and electron acceleration would draw energy from the same source (waves),
giving rise to a relationship between heating rate (hence time derivative of thermal emission) and HXRs.
Such a relationship may produce correlations similar to the Neupert effect \citep{1968Neupert},
which is observed in a significant fraction of flares 
(about half those studied by \citet{1999McTiernan} and \citet{2002Veronig}).
There is, however, an advantage to wave models, that if energy were delivered to the chromosphere (at least in part) by waves,
then thermal and HXR emission would not need be as closely related in all cases as is required by electron-beam-only models;
subsequently, flare models that include wave fluxes may be able to better explain non-Neupert flares.
Unfortunately, the exact form of relationship that would arise from a wave model,
and the circumstances under which it would hold are not known at this time.

A further point to discuss is that not all wave energy is lost before downgoing waves pass into the solar interior.
This provides a mechanism by which energy and momentum can be transferred from the corona to the interior during a solar flare,
as required to excite sunquakes (flare-generated seismic waves) \citep{1998Kosovichev,2006Kosovichev}.
Some aspects of sunquake excitation by direct wave coupling were previously presented by \citet{2012HudsonMomentum} and,
although such a connection remains speculative,
the present work shows waves with periods of several seconds have the best transmission characteristics
to enter the chromosphere and pass undamped into the solar interior.


This study has used linear models based on the preflare atmosphere to establish initial transmission and damping coefficients,
subject to a variety of simplifying assumptions that limit the validity of these models.
These have established that MHD waves are of interest to the flare chromosphere
and our results justify further investigation using more sophisticated models.
Limitations of the present models and their possible consequences on the results are now discussed.

The initial transmission and damping of waves with periods of 1 second or less is so efficient
that it will quickly lead to expansion of the heated layers, changing densities, 
increased radiative emissions and changes to the ionization state;
none of the above having been considered by the present investigation.
Thus, non-linear studies are required to properly determine longer term evolution, 
even over the few minutes of the flare impulsive phase.
Such studies must also give careful consideration to
conduction from the heated upper atmosphere to the TMR,
and optically thick radiative transfer and losses (which are sensitive to spectral lines),
effects that were avoided in the present study by the linear assumption.

During longer term evolution, we expect that chromospheric evaporation
will decrease the density ratio between the corona and chromosphere,
increasing transmission to the chromosphere beyond that found in this study.
Thus, waves may become an increasingly effective mode of energy transport as the atmosphere evolves.
At the same time, changes in density and temperature will alter collision frequencies.
In the TMR, heating will lead to increased ionization, increasing $\nu_{n,i}$, 
and the proton contribution to $\nu_{n,i}$ will also increase through its temperature dependence 
(see equations (\ref{eq:nu:Hp}) and (\ref{eq:nu:HM})).
This will act to decrease the effectiveness of wave damping by ion-neutral friction.
On the other hand, density decreases due to expansion of the heated layers will act to decrease $\nu_{n,i}$ 
and therefore increase the effectiveness of wave damping.
Electron resistivity will also be affected: increased ionization will 
increase the ion contribution and decrease the neutral contribution,
while direct temperature increases and density reductions decrease both.
The full picture is therefore likely to be a series of positive and negative feedbacks
that increase wave transmission but ultimately quench wave heating.

The present study is additionally limited by a 1D assumption.
It is desirable that this should be relaxed to consider incident waves with finite transverse wavelengths 
(which must be present given that the corona is highly structured)
and magnetic field geometry consistent with height variation of the magnetic field strength
(as discussed in Section \ref{sec:1d_model}, 
the present model allows the equilibrium magnetic field strength to vary with height, 
although horizontal components of the equilibrium field are neglected; 
this does not appear to introduce undue artifacts, however, it means that the model is not fully self-consistent).
Non-linear wave amplitudes should also be considered for flares,
leading to wave steepening, with increased dissipation, altered transmission properties, and generation of turbulence.
The last area we identify for future improvement is modeling of the different particle species and their collisions.
Ion-neutral collisions were assumed elastic in this work, 
which is reasonable when the relative ion-neutral fluid velocity is less than typical thermal speeds.
It is possible, however, that studies of the longer term 
atmospheric response to wave energy will show this condition breaks down,
in which case collisional or Alfv\'en ionization may 
play some role \citep[e.g.][and references therein]{2001LaiCIV}.
Furthermore, in a rapidly evolving atmosphere, 
populations ought to be solved explicitly in time without equilibrium assumptions.

In conclusion, this work has shown that flare-generated MHD waves can transport energy from the corona to the 
chromosphere and heat the chromosphere, with short period waves (a few seconds or less) being most effective.
They are particularly interesting as a mechanism for heating the TMR, thereby contributing to white light emissions.
This paves the way for more sophisticated modeling that may establish a viable counterpart to electron-beam flare models.

\section*{Acknowledgements}
AJBR is a research fellow of the Royal Commission for the Exhibition of 1851.
LF was supported by STFC grant ST/I001808, the EC-funded FP7 project HESPE 
(FP7-2012-SPACE-1-263086) and Leverhulme Foundation grant F00-179A.

\bibliographystyle{plainnat.bst}

\end{document}